\documentclass[12pt,a4paper]{article}
\usepackage{graphicx}
\usepackage{amsmath}
\usepackage{amssymb}
\usepackage{cite}

\begin{document}
\begin{center}
\LARGE{\bf Correlations of projectile like fragments in heavy ion reactions at Fermi energy}
\end{center}
\vspace {1cm }
\begin{center}
{\it Jayanti Rama Rao\footnote{On leave from Benares Hindu University, Varanasi, India}, H. Machner\footnote{also at Univ. Duisburg-Essen, FB Physik, Duisburg, Germany}, G. Buhr, M. Nolte, M. Palarczyk}\\
%
%
{Institut f\"{u}r Kernphysik, Forschungszentrum J\"{u}lich,
J\"{u}lich, Germany }
\
\end{center}
%
%
\begin{center}
Abstract
\end{center}
Correlations between pairs of projectile-like
fragments, emitted by the system ${^{16}O}+{^{197}Au}$ at the laboratory
bombarding energy of 515 MeV, have been studied under two
stipulated conditions: (1) at least one member of the pair is
emitted at an angle less than the grazing angle for the system, (2)
both the members of the pair are emitted at angles larger than the
grazing angle. A surprisingly large difference, by more than an
order of magnitude, is found between the correlations for the two
cases. This observation could be explained on the basis of a simple
semi-classical break up model. Further analysis of the variation of
the charge correlation function with the difference in the nuclear charges
of the correlated pair showed trends which are consistent with an "inelastic break up process", in which the projectile breaks up at
the radius of contact, in such a way that, one fragment (preferably the lighter) is emitted to one side within the grazing
angle, while the
other orbits around the target nucleus for a while and emerges on
the other side, at a negative scattering angle, much like in a deep
inelastic scattering.

\section{Introduction}

In heavy ion reactions
induced at bombarding energies below 10 MeV per nucleon the cross
section for collisions is largely found in fusion and deep
inelastic processes. In both channels, the interacting nuclei
exhibit a large degree of coherence in the sense that very few
direct third particles are emitted in the collision. Since the
macroscopic velocities are small in comparison with the intrinsic
speeds of nucleons, the propagation of disturbances through the
whole system is rather fast. This is the reason for the
manifestation of one-body mean field effects in this region.

This situation gradually changes as the bombarding energy increases. An important threshold is probably reached when the excitation energy of a nuclear system is close to its binding energy. In this case, the system largely loses its cohesion and fragmentation is more likely to set in. This limiting case is represented by an average Fermi velocity of 0.2c corresponding to a projectile energy of 20 MeV$\times$A and an intrinsic transit time of 2.5$\times 10^{-22}$s for nuclear dimensions. Up to this point, the macroscopic features of heavy ion reactions may be characterised by a variety of equilibrating processes. They range from the fully equilibrated fusion reactions, followed by deep inelastic and quasi elastic collisions with varying degrees of equilibration, to non-equilibrating direct processes. At higher energies, when the excitation energy per nucleon increases above the Fermi kinetic energy, the quanta1 nature of nucleons is expected to become less important and the one-body mean field gives way to direct two-body collisions as the main governor of nucleonic motion.

In the scenario presented above, the intermediate energy region corresponding to the Fermi energy domain of 20 - 50 MeV per nucleon, is spoken of as "transition region" \cite{Suraud89, Gross90, Moretto93, Fuchs94}, marked by qualitative changes in the heavy ion interaction characteristics. This energy also defines the threshold for the "supersonic region", where the particle velocities exceed the velocity of sound in nuclear matter.

An
important facet of interaction mechanism is the dissipation of
kinetic energy of the colliding nuclei. Experimentally this is
studied by the energy and angular distributions as well as the
correlations between the ejectiles. However, theoretically, there
seems to be a lot of confusion about this important aspect of
physics in the intermediate energy region. There are two, seemingly
equivalent but conceptually contradictory, viewpoints:
thermodynamical and dynamical. A set of models \cite{Bondorf85, Koonin87, Friedman90} based on the first viewpoint envisages the occurrence of localised zones of high excitation or "hot spots", as they are called. Formation of such
zones would imply that the dissipation is sufficiently effective to
thermalise the collective energy in the contact zone of the
colliding nuclei, whereas heat conduction is not fast enough to
spread the heat over the whole nuclear system prior to particle
evaporation. Present evidence for hot zones is essentially derived
from light particle emission with pre-equilibrium characteristics.
Their distributions are generally parameterised by Maxwellian
distribution for thermal emission from hot moving sources, and, the
presence of an "intermediate velocity source" in the
three-source-fits serves as an identification of the hot emitting
zone.

For the second set of models \cite{Moehring91, Bonasera87, Royer87} based on a contradictory
viewpoint, the light particle emission is the result of a
dynamical break up of the projectile. They attribute it
essentially to the nuclear forces which act with different strength
on different parts of the colliding projectile and thereby break
the binding between its constituents. Projectile parts coming next
to the target are strongly decelerated by the friction force which
cuts the binding to the rest. While the "trapped" part orbits
around the target for some time due to viscous drag, the broken
part proceeds forward with almost the beam velocity. The whole
reaction proceeds on a short time scale, without passing through a
phase of (partial) thermal equilibration as a necessary
intermediate step for the emission, in contradistinction to the hot
spot models.

Experimentally for all ejectiles, a clear correlation
is observed between their energy loss and angle of emission. But
this can be explained by both the viewpoints insofar as the light
particles are concerned. It has been argued that a projectile like
fragment (PLF) orbiting on the target surface, shaking off light
particles during and because of deceleration (and getting excited,
of course) is not, conceptually and phenomenologically, distinct
from a rotating hot spot.

Schwarz and collaborators \cite{Schwarz92, Terlau88} sought to clarify this situation by studying the coincident angular distributions between projectile like fragments and alpha particles. In experiments with 20 MeV$\times$A ${^{20}Ne}$ ions bombarding ${^{197}Au}$ target, they found evidence for beam velocity alpha particles, preferentially emitted to forward angles, having correlations with strongly damped PLFs, slowed down orbiting around the target to actually negative scattering angles. Complementary to this, in an experiment with 26 MeV$\times$A ${^{32}S}$ on ${^{197}Au}$, fast (more than two-thirds beam velocity ) PLFs proceeding at forward angles were found to be in coincidence with degraded alpha particles emitted at negative angles. Both these observations are projected to support a picture of dynamical break up of the projectile followed by the ( deep inelastic ) rotation of one of the fragments around the target for a while. One serious problem in resolving the ambiguity by the study of coincident angular correlations between PLFs and light particles is the sizeable and interfering contribution of such particles due to the sequential decay of the primary excited ejectiles. These events, particularly at forward angles, have to be carefully identified and eliminated to isolate and study the genuine events of direct projectile break up due to frictional forces . Experimental uncertainties in the " subtraction procedure " tend to reduce the force of the arguments. On the other hand, the coincident correlations between PLFs heavier than alpha particles are much less influenced by sequential decay. One simple way to obtain a direct signature of projectile break up vis-a-vis hot spot formation is to study the mutual correlations of all fragment pairs emitted on either side of the beam axis and to examine the trend of correlation as a function of the opening angle between the pair of fragments. It is a well established fact \cite{Boal90} that the charge correlation function is modulated, not only by the extent and life time of the source emitting the particles, but also, by the dynamics of the emission process. The latter aspect is further elaborated in the next section.

\section{Idea of the experiment}

The inspiration for the experiment came from our previous studies on the break up of loosely bound projectiles such as deuterium and ${^6Li}$ at low bombarding energies of 25-80 MeV \cite{Pampus78, Bechstedt80}. In those studies an unmistakable evidence was found for an "inelastic break up" mode, in which the projectile breaks up into two fragments, one of which comes out from the reaction zone as a quasi elastic particle while the other fuses with the target and becomes unobservable. The energy and angular distributions of the observable spectator-like fragment could be satisfactorily explained on the basis of post DWBA formalism adapted to continuum states. Such a scenario was recently found to hold also for ${^{16}O}$ induced reactions at 25 MeV$\times$A \cite{Gadioli03}.  In view of the loss of cohesion at Fermi velocities, as pointed out earlier, the heavy ion projectile in the intermediate energy region is virtually a loosely bound entity as the composite particles at low energies. Another interesting and simplifying feature of Fermi energy regime is the expectation that a classical approach is possible for the description of heavy ion collisions and the main features of the reaction can be explained by assuming a relative motion of the nuclei along classical trajectories in the field of conservative and nonconservative (frictional) forces. As was pointed out by Strutinsky \cite{Strutinsky64}, the classical approach can be applied under the condition that the reduced wave length, $\lambda/2\pi=\hbar/p$, is much shorter than the range of impact parameters, $\delta b$, that contribute to the heavy ion collision. The deep inelastic reactions represent a perfect example of the classical process, since the range of angular momenta associated with these reactions, $\delta l=\delta b\times p$, is large (of the order of 100) as compared with the Planck constant $\hbar$.

In classical collisions under purely conservative fields, the classical deflection function defines the distribution of trajectories according to the impact parameter, starting from the one at the grazing angle, $\theta_{graz}$,  corresponding to the condition, $d\theta/dl=0$, to the one at the "orbiting" angle, $\theta_0$, corresponding to $d\theta/dl=\infty$. The latter condition occurs for an orbital angular momentum, $l_{crit}$ where the nuclear, centrifugal and Coulomb forces balance one another ($dV/dr = 0$), the fission barrier disappears and, consequently, the system is poised for rotation or break up. The "grazing" condition is fulfilled for $l_{max}$ at the grazing angle, corresponding to the most peripheral collisions.

\begin{figure}[!h]
\begin{center}
\includegraphics[width=0.45\textwidth]{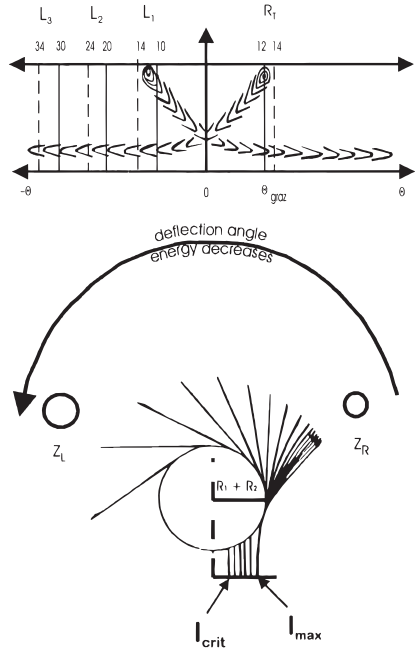}
\caption{Lower panel: Classical trajectories of a heavy ion projectile in the mean field
approximation. $R_1$ and $R_2$ are the radii of the target and
projectile. $Z_L$ and $Z_R$ are fragments due to projectile break up
caused by frictional forces at the surface (see text) . (Upper panel:
Schematic contour diagram of the cross section for collision in
energy-angle space. The solid and dashed vertical lines indicate
the angular positions of various detectors, $L_1$, $L_2$, $L_3$ and $R_T$, for
the study of fragment pair correlations (see text).}
\label{Fig1}
\end{center}
\end{figure}

The addition of a non-conservative frictional force \cite{Wilczynski73}, pushes $\theta_0$ down to negative angles, and the trajectories are deflected even to the opposite side of the grazing angle, as shown in the lower panel of Fig.\ref{Fig1}.
The upper panel of Fig. \ref{Fig1} shows the distribution
of cross section in energy-angle space. In the schematic contour
diagram, the cross section maximum is shown by the "mountain" at
the grazing angle, $\theta_{graz}$, on one side, while its "ridge" runs
all the way to (negative angles on) the other side, where the
cross section is minimum. The two symmetrical branches in this
diagram correspond to the possibility of scattering on either side
of the target nucleus. The energy distribution shows a clear
correlation with the scattering angle. For peripheral collisions, ($l=l_{max}$), the energy losses are not very large. But with
decreasing angular momentum, the trajectories are deflected towards
forward (and further to negative) angles, and simultaneously, the
energy losses increase.

At Fermi velocities in the transition region, the possibility of direct break up has to be superimposed on this picture. The nuclear friction provides the cutting sharp edge for the break up of the projectile moving at "supersonic" velocity. The approaching frontal part of the projectile is decelerated and cut off from the rest. Consequently, as shown in Fig. \ref{Fig1}, one has a beam velocity projectile fragment ($Z_R$), preferentially scattered towards the grazing angle, while the decelerated fragment ($Z_L$), dragged around the target nucleus for a while, emerges (after a "deep inelastic" scattering) from the other side or may even break up into smaller fragments.

In this scenario, the study of fragment pair correlations is interesting and illuminating. For example, with two detectors on either side of the beam axis, the correlations should show a big difference, if one of the detectors is inside or outside the grazing angle. Based on this idea, the mutual correlations of all fragment pairs of the projectile are studied in this experiment, for different angular situations of the detectors as indicated in the lower panel of Fig. \ref{Fig1}, with a view to obtain a clearer insight into the reaction mechanism at intermediate energies. This is the motivation for our work.

\section{Calculation of the grazing angle}

The grazing angle plays a crucial
role in the model presented in the previous section. Its connection to the highest angular momentum partial wave, $l=l_{max}$ participating in a
reaction and the Coulomb parameter $\eta$ can be found in \cite{Wilcke80}. It would be of course the best to take this angle from experiment. Another method is to calculate the
quarter-point angle, $\theta_{1/4}$, using the asymptotic values of the parameters $\eta$ and wave number $k$, with $R_{int}$ defined in terms of matter-half-density radii of the projectile and target. One such tabulated value \cite{Wilcke80} for the system, ${^{16}O}+{^{197}Au}$, at $E_{lab}$ = 515 MeV, was given as $\theta_{1/4}\sim 10$ degree. It depends on $R_{int}$ the interaction radius. The value in \cite{Wilcke80} agrees nicely with the one given in \cite{Roussel87}. However, it was pointed out \cite{Wilcke80}[18] that,
for strongly damped collisions, a modified value of the Coulomb parameter has to be used, as suggested by Wilczynski \cite{Wilczynski73, Galin76}, taking into
account $v'$, the relative velocity of the nuclei in the entrance
channel at the interaction radius. Using Wilczynski's prescriptions
\cite{Wilczynski73}, we have calculated the grazing angle to be about 12.5 degree for the present system. The
placement angles of various detectors in the experimental set up
were chosen based on this consideration.

\section{Experimental details}
The experiments were carried out at the J\"{u}lich Isochronous
Cyclotron. Beams of ${^{16}O^{7+}}$ ions at the laboratory energy of
515 MeV were transported to a scattering chamber to hit self-supporting gold targets. The beams were then dumped in a shielded
Faraday cup 4m downstream. The targets were of pure gold, made by
rolling, and had thicknesses of 10 mg/cm$^2$ for coincidence
measurements and 1.5 mg/cm$^2$ for singles studies. For calibration
purposes a deuterated polyethylene ($CHD$) target of thickness 12
mg/cm$^2$ was used. Charged particles were detected using silicon
counter telescopes. The thicknesses of $\Delta E$ and $E$  detectors were
chosen according to their placement angle with respect to the beam
axis.
\begin{figure}[!h]
\begin{center}
\includegraphics[width=8 cm]{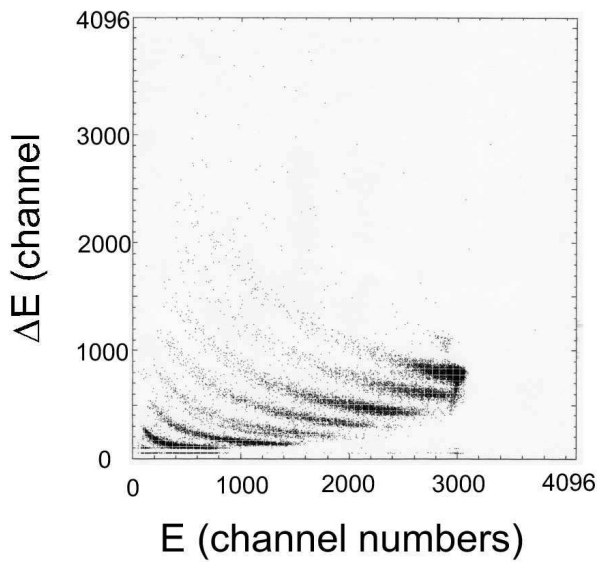}
\caption{$\Delta E-E$ scatter plot of the telescope $R_T$.}
\label{Fig:Delta-E-E_4}
\end{center}
\end{figure}
Two detector telescopes, $L_1$ and $R_T$, placed at -l0 degrees and +12 degrees with respect to the beam had $\Delta E$ detectors of thickness 200 $\mu$m and E detectors of 5 mm each and 100 mm$^2$ area each. Two other telescopes, $L_2$ and $L_3$,
placed at -20 degrees and -30 degrees with respect to the beam, had $\Delta E$ detectors of 50 $\mu$m and $E$ detectors of 2 mm each, but 150 mm$^2$ area. All the telescopes had
a veto detector of thickness 60 $\mu$m behind each of them. The
right side detector, $R_T$, served as a trigger detector for
coincidence. Its response to fragments is shown in Fig. \ref{Fig:Delta-E-E_4}. The solid angles of the telescopes, defined by apertures of Tantalum, were slightly
different but they were kept the same for singles and coincidence
measurements, so that they will cancel out in calculating the
charge correlation function. Signals of each $\Delta E$ and $E$ detectors as well as
their sum were fed through standard electronics to
analog-to-digital converters and then stored on magnetic tape in an
event-mode type of recording for later off-line analysis. The
procedure to balance the summation amplifiers and other details of
logic circuits are reported in our previous publications \cite{Machner85, Buhr92}. A typical-time spectrum for the coincidence between a pair of charged
fragments, $c_1$ and $c_2$, detected at angles, $\theta_1=12$ degrees and $\theta_2=-10$ degrees, is shown in Fig.\ref{Fig2}.
\begin{figure}[!h]
\begin{center}
\includegraphics[width=0.4\textwidth]{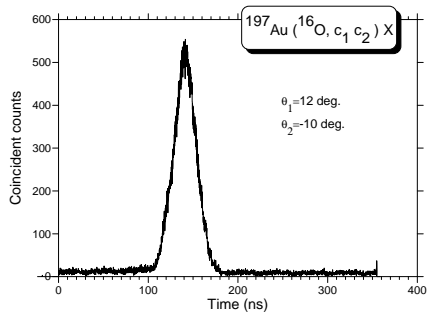}
\caption{Time spectrum for the coincidence of two charged fragments, $c_1$ and $c_2$, emitted into the detectors at 12 degrees and -10 degrees, respectively, in the reaction ${^{197}Au}({^{16}O},c_1c_2)X$ at $E_{lab}$ = 515 MeV. }
\label{Fig2}
\end{center}
\end{figure}

The experiment was conducted in two parts: In
the first part, the right side detector $R_T$ was placed at +12 degrees with
respect to the beam, so that it is inside the grazing angle of
+12.5 degrees. The left side detectors, $L_1$, $L_2$ and $L_3$ were placed
respectively at -l0 degrees, -20 degrees and -30 degrees, on the other side of the beam axis. These positions are indicated by the vertical solid lines in the upper panel of Fig. \ref{Fig1}. From the viewpoint of the "mountain" at the negative grazing angle, -12.5 degrees, the detector $L_1$ is inside, while the others, $L_2$ and $L_3$, are outside the grazing angle. In the second part of
the experiment, the right side detector $R_T$ was shifted to +14 degrees, so
that it is completely outside the grazing angle of +12.5 degrees. At the
same time the detectors, L1,L2 and L3, were shifted to -14 degrees, -24 degrees and -34 degrees, respectively, so that all of them are also outside the
grazing angle. These positions are indicated by the dashed lines in the upper panel in Fig. \ref{Fig1}.

By switching a simple AND/OR gate, the singles spectra in
all the four detectors, $L_1$, $L_2$, $L_3$ and $R_T$, and, coincidence
spectra between the three pairs of left and right detectors, $L_1R_T$,$L_2R_T$ and $L_3R_T$, were recorded under identical geometrical
conditions in different runs. The singles spectra of the PLFs with
$Z=2$ to $Z = 7$, were sorted out from the $\Delta E-E$ matrix of each
detector using "banana" shaped gates which included all the isotopes of a
given $Z$ except for nitrogen. Since the elastic line of $^{16}O$ is leaking into the nitrogen band (see Fig. \ref{Fig:Delta-E-E_4}) that area has been excluded. This leads to the sharp cut off at 515 MeV in the nitrogen singles spectrum. The spikes in some of the spectra are most probably due to the limoted statistics. The coincidence spectrum of a given ordered pair of PLFs
was obtained by sorting out the coincident $\Delta E-E$ matrices of the
left and right detectors with appropriate banana gates
corresponding to the respective PLFs. Some typical spectra are
shown in Figs.3-4.
\begin{figure}[!h]
\begin{center}
\includegraphics[width=0.45\textwidth]{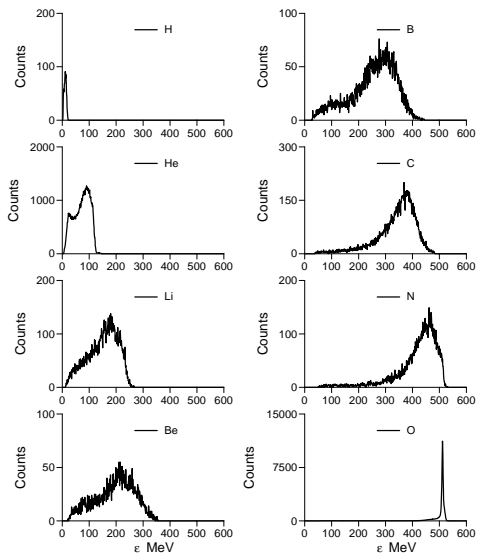}
\caption{Singles spectra of projectile like fragments and elastically scattered oxygen at 12 degrees (lab. system). }
\label{Fig3}
\end{center}
\end{figure}
\begin{figure}[!h]
\begin{center}
\includegraphics[width=0.45\textwidth]{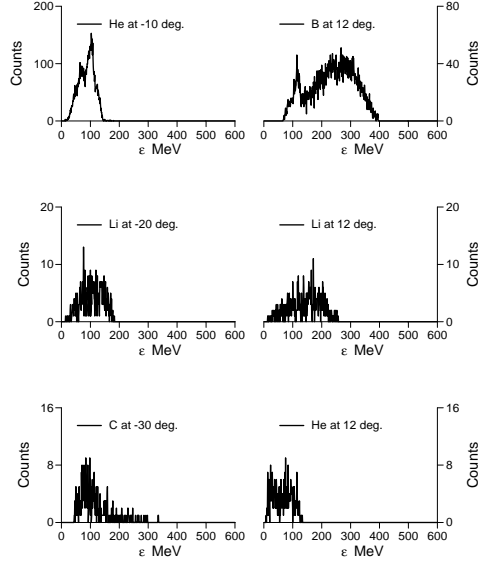}
\caption{Typical coincidence spectra (lab. system) of
fragment pairs, $He-B$, $Li-Li$ and $C-He$, detected in the left and
right detectors, at the angles noted in the figure.}
\label{Fig4}
\end{center}
\end{figure}

The fragment spectra were then converted into
double differential cross sections using standard formulae. Singles
cross section is given by
\begin{equation}
\frac{d^2\sigma}{d\Omega d\epsilon} = \frac{Z_PeA_TN_s\sin\phi 10^{24}}{Q\rho L\Delta\Omega \Delta\epsilon (1-\tau/100)}\,\, \left(\frac{\text{barn}}{\text{sr MeV}}\right)
\end{equation}

where
\begin{flushleft}
\begin{tabular}{ll}
$Z_P$ & Atomic number of the projectile \\
$e$ & elementary charge \\
$A_T$ & Atomic mass of the target (g/mol) \\
$N_S$ & number of counts in the energy bin of $\Delta\epsilon$ \\
 & in the singles spectrum \\
$\phi$ & Angle between target plane and beam direction \\
$Q$ & Charge collected in the Faraday cup (Cb) \\
$\rho$ & target thickness (g/cm$^2$) \\
L & Avogadro number (1/mol) \\
$\Delta \Omega$ & solid angle of the detector (sr) \\
$\Delta \epsilon$ & Width of the energy bin in the spectrum (MeV) \\
$\tau$ & dead time (percent) \\
\end{tabular}
\end{flushleft}

The coincidence cross section is given by

\begin{gather}
\frac{d^4\sigma}{d\Omega_1 d\epsilon_1d\Omega_2 d\epsilon_2} = \frac{Z_PeA_TN_{coi}\sin\phi 10^{24}}{Q\rho L\Delta\Omega_1 \Delta\epsilon_1
\Delta\Omega_2 \Delta\epsilon_2 (1-\tau/100)} \nonumber \\
\left(\frac{\text{barn}}{\text{sr$^2$ MeV$^2$}}\right)
\end{gather}
where the indices 1 and 2 refer to the PLFs
detected in the left and right detectors, and $N_{coi}$ is the
corresponding number of coincidence counts for the PLF pair.

The
systematic errors in the cross section measurement are: 5$\%$ error in
the charge collected in the Faraday cup, 1$\%$ error in the target
thickness, 5$\%$ error in dead time and 2$\%$ error in each of the solid
angles (which however cancels out in the calculation of the
charge correlation function). In addition, statistical errors in the
singles and coincidence counting rates are taken into account for
each individual case. The energy dependent Correlation function, $R(c_1,c_2)$, is usually defined in terms of the differential cross
sections as
\begin{equation}
R(\epsilon_1,\epsilon_2)+1 = \sigma_0\frac{d^4\sigma (\epsilon_1\epsilon_2)}{d^2\sigma(\epsilon_1)d^2\sigma(\epsilon_2)}
\end{equation}
where $\sigma_0$ is a constant, which involves the multiplicities of the PLFs and the total reaction cross section.

\section{Results and discussion}

\begin{figure}[!h]
\begin{center}
\includegraphics[width=0.45\textwidth]{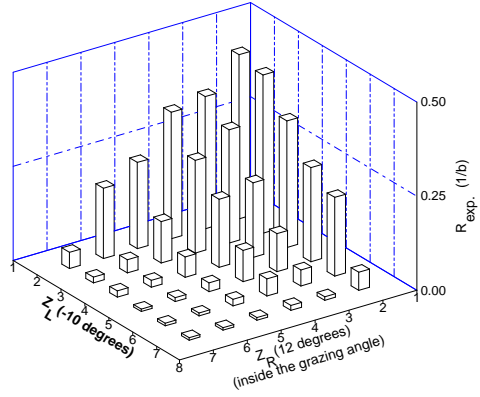}
\caption{Three dimensional viewgraph showing the observed correlation, $R_{exp}$, as function of the atomic numbers, $Z_L$ and $Z_R$, of the PLFs
detected in the left side detector $L_1$ at -l0 degrees and the right side
detector $R_T$ at 12 degrees. $R_T$ is inside the grazing angle of 12.5 degrees for the present case.}
\label{Fig5}
\end{center}
\end{figure}

\begin{figure}[!h]
\begin{center}
\includegraphics[width=0.45\textwidth]{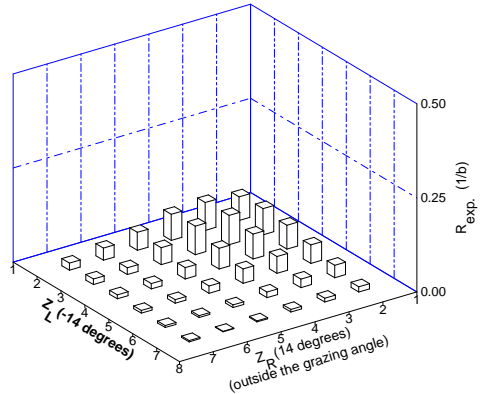}
\caption{Same as Fig. \ref{Fig5}, except that $L_1$ is at
-14 degrees and $R_T$ is at 14 degrees. $R_T$ is outside the grazing angle of 12.5 degrees.}
\label{Fig6}
\end{center}
\end{figure}

\begin{figure}[!h]
\begin{center}
\includegraphics[width=0.45\textwidth]{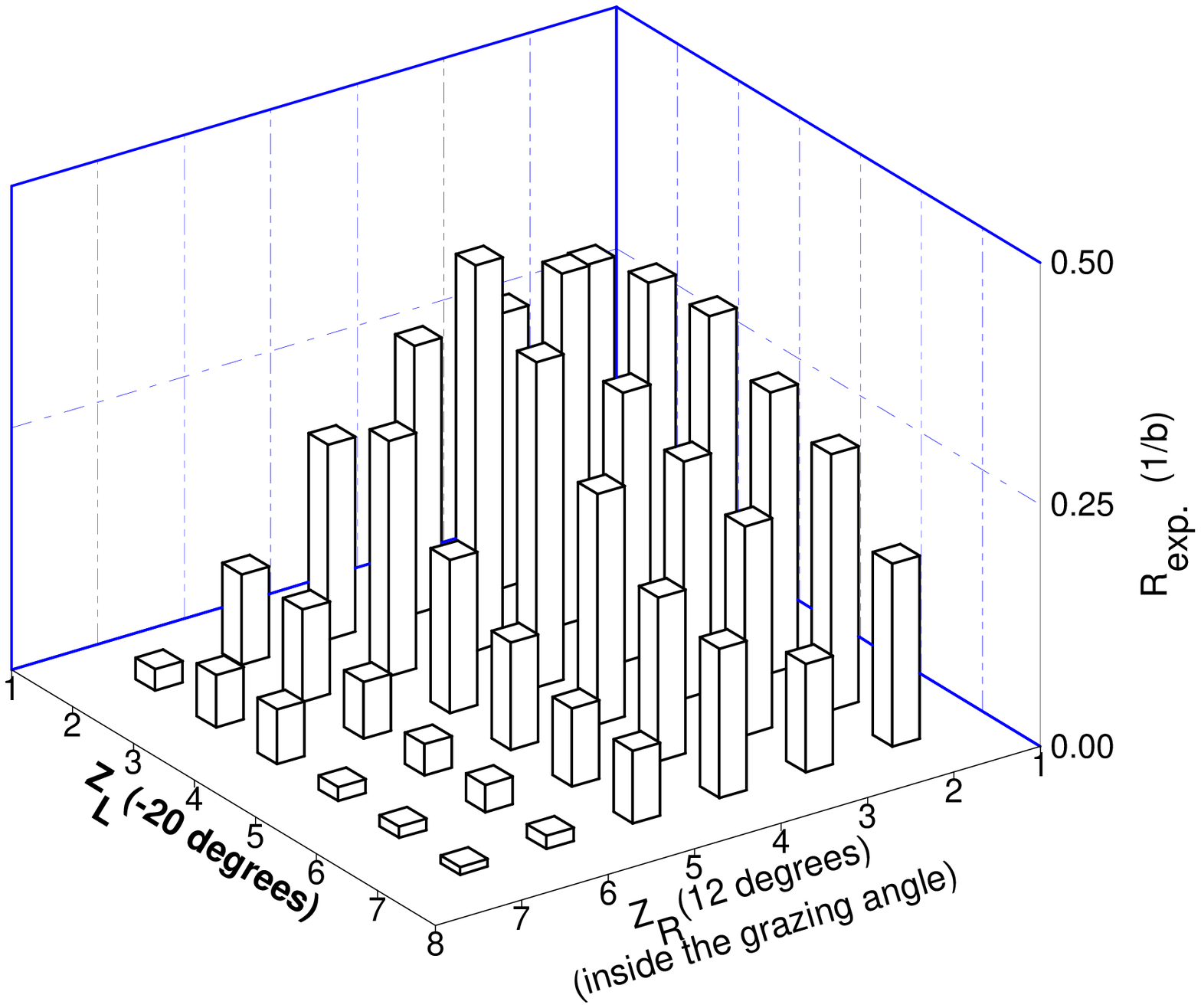}
\caption{Same as in Fig. \ref{Fig5}, except that $L_2$ is at -20 degrees. $R_T$ is inside the grazing angle.}
\label{Fig7}
\end{center}
\end{figure}

\begin{figure}[!h]
\begin{center}
\includegraphics[width=0.45\textwidth]{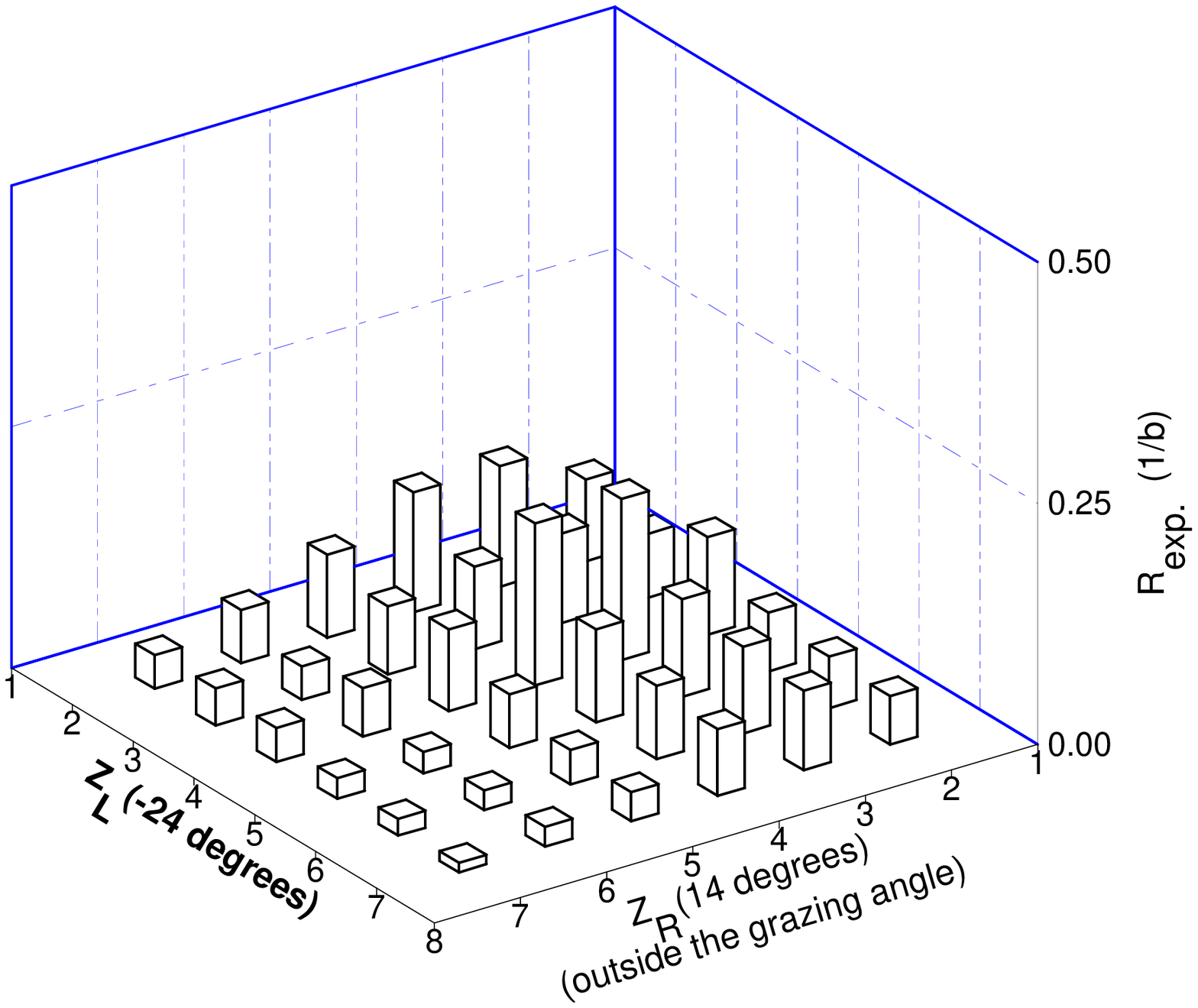}
\caption{Same as in Fig. \ref{Fig6}, except that $L_2$ is at
-24 degrees. $R_T$ is outside the grazing angle.}
\label{Fig8}
\end{center}
\end{figure}

\begin{figure}[!h]
\begin{center}
\includegraphics[width=0.45\textwidth]{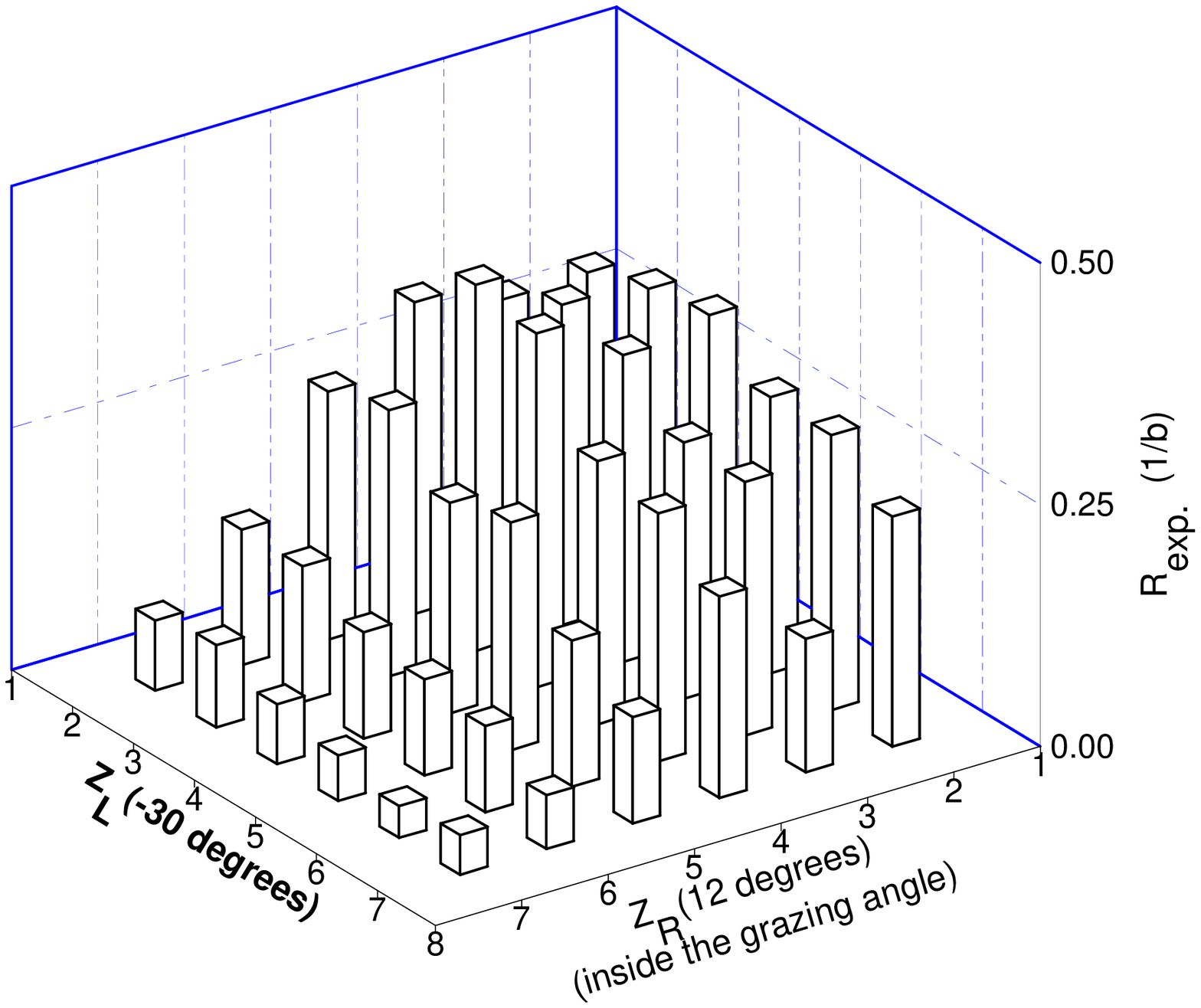}
\caption{Same as in Fig. \ref{Fig5}, except that $L_3$ is at -30 degrees. $R_T$ is inside the grazing angle.}
\label{Fig9}
\end{center}
\end{figure}

\begin{figure}[!h]
\begin{center}
\includegraphics[width=0.45\textwidth]{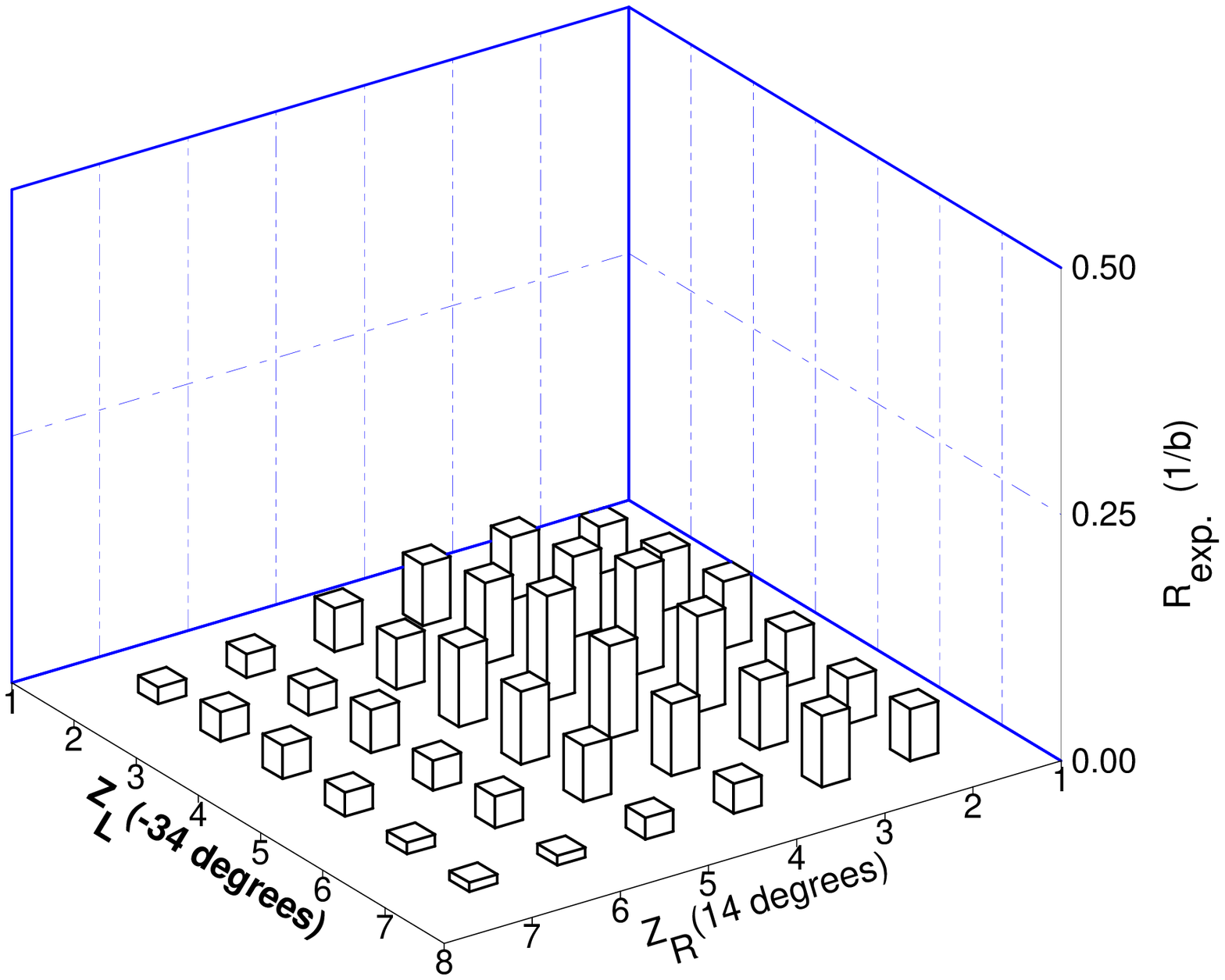}
\caption{Same as in Fig. \ref{Fig6}, except that $L_3$ is at -34 degrees. $R_T$ is outside the grazing angle.}
\label{Fig10}
\end{center}
\end{figure}

For the purpose
of studying the correlations in the present experiment, the
measured double differential cross sections are integrated over the
fragment energy $\epsilon$, and redesignated as $d\sigma_{12}$for the coincidence cross section and $d\sigma_1$ and $d\sigma_2$  for the singles cross sections of the pair of fragments with atomic numbers $Z_1$ and $Z_2$. The ratio of
cross sections as defined below, $R_{exp}$, usually termed as the
experimental correlation, contains a Charge Correlation
function, $R_Z$, given by
\begin{equation}
R_{exp} = \frac{R_Z+1}{\sigma_0} =\frac{d\sigma_{12}}{d\sigma_1d\sigma_2}.
\end{equation}
The values of R are presented as three
dimensional graphs  in Figs.\ref{Fig5}-\ref{Fig10}, for various angular placements of the detectors, as a function of the atomic numbers $Z_L$ and $Z_R$ of
the PLFs detected in the left and right detectors respectively.
It can be seen from these figures that whenever the right side
detector is at the angle 12 degrees, the values of $R_{exp}$ are quite large,
but when this angle is just changed to 14 degrees, the values are
diminished very substantially. This observation is true for all
angular positions of the left side detector from -10 degrees to -34 degrees. The
sudden vanishing of the correlations for a change from 12 degrees to 14 degrees is quite significant in view of the fact that the calculated
grazing angle for the present system is about 12.5 degrees as mentioned
previously. At 12 degrees the right side detector is \emph{inside} the grazing angle and for 14 degrees it is \emph{outside} the grazing angle, even after
taking into account the angular resolution of the detector, which
is about 0.8 degrees. The big difference observed in the correlations is
consistent with the expectations on the basis of the semi-classical
model described in Section 2.

\begin{figure}
\begin{center}
\includegraphics[width=0.45\textwidth]{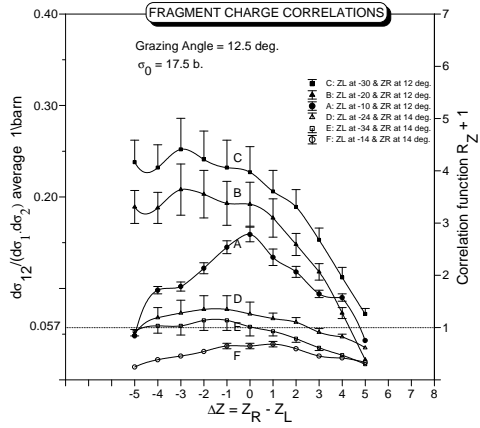}
\caption{\label{Fig11}Trends of correlation as a function of $\Delta Z$, the difference in the nuclear charges of the pair of projectile like fragments (see text). The curves are drawn only to guide the eye.}
\end{center}
\end{figure}

In order to get a quantitative idea of the
correlations, we studied the variation of $R_{exp}$ as a function of $\Delta Z$, for an ordered pair (L,R) of PLFs, where $\Delta Z= Z_R - Z_L$ is the
difference of their atomic numbers $Z_L$ and $Z_R$, and they are
detected in the left and right detectors respectively. Since PLFs
with $Z = 2$ to 7 are detected on both sides in the present
experiment, $\Delta Z$ varies from -5 to +5, for all possible pairs. The
average value $<R_{exp}>\; =\; <d \sigma_{12}/d\sigma_1 d\sigma_2>_{average}$ is calculated
for each value of $\Delta Z$, and plotted in Fig. \ref{Fig11},
for the six angular
situations of the detectors. It can be seen that the three bottom
curves, (D,E,F), corresponding to the right side detector at 14 degrees (outside the grazing angle), show minimum correlation, which is
taken to correspond to $R_Z = 1$, as shown by the dashed line at $<R_{exp}>
= 0.057$. From this a value of 17.5 b for the normalisation
constant, $\sigma_0$, is deduced.

The extracted values of the Correlation
function $R_Z$ are shown in the same \ref{Fig11}, on the right hand side of
the y-axis. It is noteworthy that the present values of $R_Z$ for wide
angle correlations of PLFs are of the same order of magnitude as
those of $R_q$ the momentum dependent Correlation function,
observed in our previous study \cite{Machner92} of small angle correlations
of alpha particles in the ${^4{He}}+{^{58}Ni}$ system at 120 MeV. In the
present experiment, several interesting new features are observed
in the trends of the Correlation function, which can be understood
on the basis of the semi-classical model presented in Section 2:
\begin{enumerate}
\item
The position of the right side detector, $R_T$, relative to the
grazing angle, ( $\theta_{graz} = 12.5$ degrees), appears to act like an on/off switch, to put on and put off correlations at 12 degrees and 14 degrees respectively, irrespective of the position of the left side
detectors. This behaviour can be understood by referring to Fig.\ref{Fig1}.
If one imagines the PLF $Z_R$ to be detected at the "mountain", near
+$\theta_{graz}$, and the other PLF $ZL$ to be on the left side, either at
-l0 degrees, or -20 degrees, or -30 degrees, on the "ridge", then correlations always exist between them, because they originate from the interaction on
the same side of the target as shown in the lower panel of Fig. \ref{Fig1}. But, if the right side detector is at 14 degrees and the left side
detectors at -14 degrees, -24 degrees and -34 degrees as shown by the dashed lines in
the upper panel of Fig. \ref{Fig1}, then, not only they are on the "opposite ridges", but more importantly, the PLFs detected by all the
detectors are mainly from statistical emission from the composite
system formed by lower angular momentum partial waves. Hence there
is minimum or no correlation between the ejectiles.
\item
The value
of the Correlation function $R_Z$ seems to increase with the negative
angle of scattering of the left side PLF, as can be seen from the
curves A for -10 degrees, B for -20 degrees and C for -30 degrees in \ref{Fig11}. This trend
is again quite consistent with the model depicted in Fig. \ref{Fig1} and
discussed in Section 2. It can be explained as follows: With
increasing relative angle, the background contribution from the
uncorrelated statistical emission decreases very fast, thus leaving
only the real correlated events to dominate.
\item
It is curious that, while the curve A in \ref{Fig11} corresponding to $Z_L$ at -10 degrees and $Z_R$ at 12 degrees is symmetrical about $\Delta Z=0$, the curves B and C corresponding
to $Z_L$ at -20 degrees and -30 degrees, respectively, are asymmetric. Further, the curves B and C are rather constant for negative $\Delta_Z$, that is, when
$Z_L$ detected on the left side is larger than $Z_R$ detected on the
right side. This indication lends support to our idea of inelastic
break up, in which a small fragment of the projectile breaks away,
without much energy loss, preferentially towards the grazing angle,
while the rest of the projectile undergoes a large dissipation of
energy in a kind of deep inelastic scattering around the target
nucleus. This is a new reaction mode, which is clearly distinct
from the "free" or elastic break up (with no energy dissipation)
observed at higher energies, as well as from the so called "incomplete fusion" observed at lower energies.
\end{enumerate}

In all cases shown in Figs \ref{Fig5} to \ref{Fig10} one can clearly see events in which sum of the atomic numbers of two projectile-like fragments is larger than the charge of the oxygen projectile. It is very likely that these are the events where $^{16}O$ projectile breaks into two parts - one which continue to move with almost the projectile velocity, and the second, which while orbiting around the target nucleus is "receiving" some matter from the target and is emerging with higher nuclear charge than originally having. The longer is the orbiting time (i.e. larger the absolute value of $\theta_2$), probability for this kind of events is increasing which is also seen when comparing e.g. Figs. \ref{Fig5}, \ref{Fig7} and \ref{Fig9}. This could in part also explain a difference in shapes of curves A, B and C seen in Fig. \ref{Fig11}.

A series of studies of the present system $^{16}O+^{197}Au$ at bombarding energies comparable to the nucleonic Fermi energy have bee performed in the past. They mostly concentrated on fragmentation into $\alpha$ particles or $\alpha$ cluster nuclei \cite{Moehring91, Harmon90, Kelly97}. This seems to be a to limiting view. From the two dimensional charge correlation functions it is evident that for instance correlations with lithium are as abundant as with helium. Pouliot et al. \cite{Pouliot91} measured mainly light-light and heavy-light coincidences including channels up to five light fragments. Their main interest are those channels where the sum of all fragments is the projectile. The present work is complementary to these studies. Previous studies found strong correlation between nucleons and projectile fragments \cite{Roussel00}, with the projectile emission angle within the grazing angle. On the whole, the
presently observed fragment--fragment correlations in the highly
asymmetric ${^{16}O}+{^{197}Au}$ system, indicate the persistence of the
binary reaction dynamics, with slight modifications, at the onset
of intermediate energy regime. More exclusive and
detailed studies of fragment--fragment correlations are needed for a
clear understanding of the reaction mechanism in the intermediate
energy region.

\section{Summary and conclusions}

In summary, the trends of
fragment--fragment correlations observed in the present experiment
support the idea of an "inelastic break up process", in which the
projectile breaks up at the radius of contact, in such a way that,
one fragment (preferably the lighter) is emitted to one side
within the grazing angle, while the second fragment orbits around
the target nucleus for a while and emerges on the other side, at a
negative scattering angle, much like in a deep inelastic
scattering. Such a model nicely fits into the evolutionary nature
of the transition region, and, into the sequence of gradually
changing interaction characteristics from one-body mean field
effects to two-body collision dynamics.
\begin{center}
Acknowledgement
\end{center}
The authors wish to thank
the Cyclotron staff of JULIC for their nice cooperation. One of the authors (JRR) is grateful to the Internationales B\"{u}ro, BMFT,
for supporting his stay in J\"{u}lich.


\begin{thebibliography}{99}
%

\bibitem{Suraud89} E. Suraud, Ch. Gregoire, B. Tamain, Progr. Part. Nucl. Phys. \textbf{23}, 357 (1989).

\bibitem{Gross90} D.H.E. Gross, Rep. Progr. Phys. \textbf{53}, 605(1990).

\bibitem{Moretto93} L.G. Moretto, G.J. Wozniak, Annu. Rev. Nucl. Part. Sci. \textbf{43}, 379 (1993).

\bibitem{Fuchs94} H. Fuchs, K. Moehring, Rep. Progr. Phys. \textbf{57}, 231 (1994).

\bibitem{Bondorf85} J.P. Bondorf, R. Donangelo, I.N. Mishustin, H. Schulz, Nucl. Phys. \textbf{A 444}, 460 (1985).

\bibitem{Koonin87} S.E. Koonin, J. Randrup, Nucl. Phys. \textbf{A474}, 173 (1987).

\bibitem{Friedman90} W.A. Friedman, Phys. Rev. \textbf{C 42}, 667 (1990).

\bibitem{Moehring91} K. Moehring, T. Srokowski, D.H.E. Gross, Nucl. Phys. \textbf{A 533}, 333 (1991).

\bibitem{Bonasera87}A. Bonasera, A., M. Di Toro, C. Gregoire, Nucl. Phys. \textbf{A 463}, 653 (1987).

\bibitem{Royer87} G. Royer, Y. Raffray, A. Oubahadou, B. Remaud, Nucl. Phys. \textbf{A 466}, 139 (1987).

\bibitem{Schwarz92} C. Schwarz, H. Fuchs, H. Homeyer, K. Moehring, A. Siwek, A. Sourell, W. Terlau, A. Budzanowski, Phys. Lett. \textbf{B 279}, 223 (1992).

\bibitem{Terlau88} W. Terlau, M. Buergel, A. Budzanowski, H. Fuchs, H. Homeyer, G. Roeschert, J. Uckert, R. Vogel, Z. Phys. A - Atoms and Nuclei \textbf{330}, 303 (1988).

\bibitem{Boal90} D.H. Boal, C.K. Gelbke, B.K. Jennings, Rev. Mod. Phys. \textbf{62}, 553 (1990).

\bibitem{Pampus78} J. Pampus, J. Bisplinghoff, J. Ernst,
T.  Mayer-Kuckuk, J. Rama Rao, G. Baur, F. Roesel, D. Trautmann, Nucl. Phys. \textbf{A 311}, 141 (1978).

\bibitem{Bechstedt80} U. Bechstedt, H. Machner, G. Baur, G., R. Shyam, C. Alderliesten, O. Bousshid, A. Djaloeis, P. Jahn, C. Mayer-Boericke, F. Roesel, D. Trautmann, Nucl. Phys. \textbf{A 343}, 221 (1980).

\bibitem{Gadioli03} E. Gadioli \textit{et al.}, Nucl. Phys. \textbf{A 708}, 391 (2003).

\bibitem{Strutinsky64} V.M. Strutinsky, ZETP (USSR)
\textbf{46}, 2078 (1964).

\bibitem{Wilczynski73}  J. Wilczynski, Nucl. Phys. \textbf{A 216}, 386 (1973).

\bibitem{Wilcke80} W. W. Wilcke, J.R. Birkelund, H.J. Wollersheim, A.D. Hoover, J.R. Huizenga, W.U. Schroeder, L.E. Tubbs, Atom. Nucl. Data Tables \textbf{25}, 389 (1980).

\bibitem{Roussel87} P. Roussel et al., Nucl. Phys. \textbf{A 477},345 (1988).

\bibitem{Galin76} J. Galin, J. Phys. \textbf{G5}, 83 (1976).

\bibitem{Machner85} H. Machner, D. Protic, G. Riepe, H.G. Bohlen, H. Fuchs, Phys. Rev. \textbf{C 31}, 443 (1985).

\bibitem{Buhr92} G. Buhr, H. Machner, M. Nolte, M. Palarczyk, J. Rama Rao, Phys. Rev. \textbf{C 45}, 705 (1992).

\bibitem{Machner92} H. Machner, M. Palarczyk, H.W. Wilschut, M. Nolte, E.E. Koldenhof, Phys. Lett. \textbf{B 280},
16 (1992).

\bibitem{Harmon90} B. A. Harmon et al., Phys. Lett. \textbf{B 235}, 234 (1990).

\bibitem{Kelly97} D. O'Kelly et al., Phys. Lett. \textbf{B 393}, 301 (1997).

\bibitem{Pouliot91} J. Pouliot et al., Phys. Rev. \textbf{C 43}, 735 (1991).

\bibitem{Roussel00} P. Roussel \textit{et al.}, J. Phys. G. \textbf{26}, 1641 (2000).

\end{thebibliography}
\end{document}